\documentstyle[floats,twocolumn,aps,prbbib]{revtex}
\include{psfig}

\newcommand{\eref}[1]{Eq.~(\ref{#1})} \newcommand{\fref}[1]{Fig.~\ref{#1}}
\newcommand{\tref}[1]{Table~\ref{#1}}
\newcommand{\ocite}[1]{Ref.\protect\onlinecite{#1}}
\newcommand{\citep}[1]{\protect\cite{#1}} \newcommand{\ie}{{\it i.e. }}

\newcommand{\xc}{{\rm xc}} \newcommand{\nl}{{\rm nl}}
\newcommand{\vdw}{{\rm vdW}} \newcommand{\vv}[1]{{\bf #1}}
\newcommand{\ii}[1]{\int{\! d#1 \,}} \newcommand{\iiab}[3]{\int_{#1}^{#2}
\! \! d#3 \,} \newcommand{\iiabf}[4]{\int_{#1}^{#2} \! \frac{d#3}{#4} \,}
\newcommand{\norm}[1]{\left|#1\right|} \newcommand{\trace}[1]{{\rm
Tr}\left[#1\right]} \newcommand{\real}[1]{{\rm Re}\left[#1\right]}
\newcommand{\imag}[1]{{\rm Im}\left[#1\right]}  \newcommand{\M}[2]{\left(\begin{array}{#1} #2
\end{array}\right)} \newcommand{\ordo}[1]{{\cal O}(#1)}
\newcommand{\tntext}[2]{\small{\tablenotemark[#1]#2}}

\begin{document}
\draft

\title{ Tractable non-local correlation density functionals for flat
surfaces and slabs }

\author{Henrik Rydberg and Bengt I. Lundqvist} \address{ Department of
Applied Physics, Chalmers University of Technology and G\"{o}teborg
University, S-412 96 G\"{o}teborg, Sweden }

\author{David C. Langreth and Maxime Dion} \address{ Center for Materials
Theory, Department of Physics and Astronomy, Rutgers University,
Piscataway, New Jersey 08854-8019 }


\maketitle

\begin{abstract}
A systematic approach for the construction of a density functional for van
der Waals interactions that also accounts for saturation effects is
described, {\it i.e.} one that is applicable at short distances.  A very
efficient method to calculate the resulting expressions in the case of flat
surfaces, a method leading to an order reduction in computational
complexity, is presented.  Results for the interaction of two parallel
jellium slabs are shown to agree with those of a recent RPA calculation
(J.F. Dobson and J. Wang, Phys. Rev. Lett. {\bf 82}, 2123 1999).  The
method is easy to use; its input consists of the electron density of the
system, and we show that it can be successfully approximated by the
electron densities of the interacting fragments.  Results for the surface
correlation energy of jellium compare very well with those of other
studies.  The correlation-interaction energy between two parallel jellia is
calculated for all separations $d$, and substantial saturation effects are
predicted.
\end{abstract}


\vspace{-0.8cm}

\section{Introduction}

The density-functional theory (DFT),\cite{HoKo64} with its local-density
(LDA)\cite{KoSh65,GuLu76} and semilocal generalized-gradient approximations
(GGA),\cite{LaMe81,Peetal92,PeBuEr96,HaHaNo99} is not only successful in
numerous applications to individual molecules and dense solids. It is also
under intense development, for instance, in order to include non-local
effects, such as van der Waals (vdW)
forces.\cite{LuAnShChLa95,AnLaLu96,DoDi96,HuAnLuLa96,KoMeMa97,AnHuApLaLu98,DoWa98,HuRyLuLa98}
The latter are needed in order to allow DFT to describe sparse matter. A
unified treatment of vdW forces at large and asymptotic molecular
separations is available,\cite{HuRyLuLa98} and a description at short
distances and at overlap is striven for.  An accurate calculation for the
interaction of two He atoms has been given,\cite{KoMeMa97} and recently the
first microscopic (RPA) calculation of the vdW interaction between two
self-consistent jellium slabs has been reported and given a density
functional (DF) account.\cite{DoWa98} The ultimate challenge is to
construct an approximate vdW DF that is generally applicable, efficient,
and accurate.

We here propose an explicit form for the vdW DF that applies to flat
surfaces, test it successfully against these slab results, and apply it to
two parallel flat semi-infinite metal surfaces. This is a case with
relevance for many physical situations, including wetting and atomic-force
microscopy (AFM). Compared to the vdW-DF approximation proposed by Dobson
and Wang,\cite{DoWa98} the virtues of our functional are the computational
simplifications gained from choosing a particular sub-class of response
functions, utilizing a differential formulation and sparse matrices, and
recognizing the insensitivity to the details of the density profiles,
simplifications which might transfer even to three dimensions.

The ubiquitous van der Waals force plays an important role for numerous
physical, chemical, and biological systems, such as
physisorption,\cite{AnWiPe88,AnPeHa96} vdW complexes,\cite{BuFoHu88} vdW
bonds in crystals, liquids, adhesion, and soft condensed matter (e.g.,
biomacromolecules, biosurfaces, polymers, and
membranes).\cite{BuFoHu88,ChSz94}

The DFT expresses the ground-state energy of an interacting system in an
external potential $v(\vv{r})$ as a functional $E[n]$ of the particle
density $n(\vv{r})$, which has its minimum at the true ground-state
density.\cite{HoKo64} The Kohn-Sham form of the functional makes the scheme
a tractable one, as it leads to equations of one-electron type, and
accounts for the intricate interactions among the electrons with an
exchange-correlation (XC) functional $E_\xc[n]$.\cite{KoSh65} This
XC-energy functional can be expressed exactly as an integral over a
coupling constant ($\lambda$),\cite{LaPe75,GuLu76,LaPe77} the so-called
adiabatic connection formula (ACF),
\begin{equation}
\label{eq:XACF}
E_\xc = -\iiab{0}{1}{\lambda} \iiabf{0}{\infty}{u}{2 \pi}
\trace{\chi(\lambda,iu) V} - E_{\rm self},
\end{equation}
where $V(\vv{r},\vv{r'}) = 1/\norm{\vv{r}-\vv{r'}}$ and where the
density-density correlation function is denoted by
$\chi(\vv{r},\vv{r'},iu;\lambda)$.\cite{FootNotation} $E_{\rm self}$ is the
Coulomb self-energy of all electrons, which is exactly cancelled by a
corresponding term in $\chi(\lambda,iu) V$. Equation (\ref{eq:XACF}) shows
a truly non-local XC interaction and is a starting point for approximate
treatments, local (LDA), semilocal (GGA) and non-local ones.

The LDA and GGA are completely unable to express the vdW interaction in a
physically sound way. The exact XC energy functional, on the other hand, of
course encompasses such interactions.\cite{LuAnShChLa95} The basic problem
of making DFT a working application tool also for sparse matter is to
express the truly non-local vdW interactions between the electrons in the
form of a simple, physical, and tractable DF.  Equation (\ref{eq:XACF}) is
then the starting point. Along these lines, we have proposed extensions of
the DFT to include van der Waals
interactions,\cite{LuAnShChLa95,AnLaLu96,HuAnLuLa96,AnHuApLaLu98,HuRyLuLa98}
with very promising results for the interaction between two atoms or
molecules, an atom and a surface, and two parallel surfaces, respectively.
This now unified approach\cite{HuRyLuLa98} applies for separated systems,
\ie when the electrons of the interacting fragments have negligible
overlaps.

The corresponding asymptotic expressions have singular behaviors at short
separations $d$. Yet one knows that the vdW forces are finite.\cite{MaNi76}
They should go smoothly over to the XC forces that apply in the interior of
each electron system. This phenomenon is often called damping, or
\emph{saturation}.\cite{TaTo84,HoAp84,NoHoHa85} Approximate saturation
functions have been proposed, in particular for the cases of vdW
molecules\cite{TaTo84} and physisorbed particles.\cite{HoAp84,NoHoHa85}

The key difficulty in extracting the vdW DF from \eref{eq:XACF} is the
computational complexity. A direct solution gives simply too many
operations on the computer. The guideline for our reduction of the number
of such operations is to exploit analytical advantages of RPA-like
approximations, to focus on the key quantity, to recast the integral
formulation into a differential one, leading to a sparse-matrix
computation, and to make maximal use of symmetry.

Our exploratory study here concerns cases with vdW forces between two flat
parallel model systems.  We first test our approximate functional on the
model system of two self-consistent jellium slabs, utilizing the recent RPA
results,\cite{DoWa98} which gives the size of the correlation-interaction
energy per unit area, showing saturation.  We then test our DF against
accurate calculations of the surface correlation
energy,\cite{PeKu99,PiEg98} showing an excellent agreement. After these
successful tests we make predictions on two parallel semi-infinite jellia.

\section{General Theory}

The $\lambda$ integration of \eref{eq:XACF} can be performed analytically
in some cases, such as in the random-phase approximation (RPA). In 1957
Gell-Mann and Bruckner (GMB)\cite{GeBr57} presented the RPA correlation
energy as a selected summation of ring diagrams, which gives a logarithmic
form.\cite{Pi63} Their study concerns the \emph{homogeneous} electron gas,
where equations simplify thanks to the three-dimensional translational
invariance and plane waves. Here we treat systems with less symmetry.

By virtue of the fluctuation-dissipation theorem, the density-density
correlation function $\chi$ is equal to the density change $\delta n$
induced by an {\it external} potential $\Phi_{\rm ext}$, \ie $\delta n=
\chi\Phi_{\rm ext}$. It satisfies
\begin{equation}
\label{eq_rpa}
\chi(\lambda,iu) = \tilde{\chi}(iu) + \lambda \tilde\chi(iu) V
\chi(\lambda,iu),
\end{equation}
where $\tilde{\chi}$ is the density response to a fully screened potential
$\Phi$, i.e. $\delta n= \tilde{\chi}\Phi$. We assume here that the coupling
dependence of $\tilde\chi$ can be neglected, when performing the $\lambda$
integral in \eref{eq:XACF}. This is true in the random-phase approximation,
where $\tilde{\chi}$ is the density response function for $\lambda=0$, and
is also true for the approximate dielectric functions, which we use
here. Equation~(\ref{eq:XACF}) then becomes
\begin{equation}
\label{eq:XACF_RPA}
E_\xc = \iiabf{0}{\infty}{u}{2 \pi} \real{\trace{\log(1-\tilde{\chi}(iu)
V)}} - E_{\rm self},
\end{equation}
where the real part means the principal branch.

To simplify this expression and to get a functional in terms of the
electron density $n(\vv{r})$, we have to focus on the key target, the
non-local part, introduce key quantities, and rewrite the expressions, in
order to make physically sound and computationally efficient
approximations. It is more convenient to introduce the polarizability or
dielectric function instead of $\tilde \chi$. The polarizability
$\vv{\alpha}$ (a matrix in the spatial positions) is defined by the
relation $\vv{P}=\vv{\alpha}\cdot \vv{E}$, where $\vv{P}$ is the
polarization. We have
\begin{equation}
\delta n = -\nabla\cdot\vv{P} = -\nabla\cdot\vv{\alpha}\cdot\vv{E} =
\nabla\cdot\vv{\alpha}\cdot\nabla\Phi,
\end{equation}
so that from the definition of $\tilde \chi$, one has $\tilde{\chi} =
\nabla\cdot \vv{\alpha} \cdot \nabla$. In turn the dielectric function is
given by $\vv{\epsilon} \equiv \vv{1}+4\pi \vv{\alpha}$. In terms of
$\epsilon$ \eref{eq:XACF_RPA} then transforms to
\begin{equation}
\label{eq:XACF_POL}
E_\xc = \iiabf{0}{\infty}{u}{2 \pi} \real{\trace{\log(\nabla\cdot
\vv{\epsilon} \cdot\nabla G)}} - E_{\rm self},
\end{equation}
where we have introduced the Coulomb Green's function $G=-V/4 \pi$ and used
$\nabla^2 G = 1$. The $\trace{\log}$ expression gives great advantage for
the further analytical and numerical treatment. The only approximation made
so far is the neglect of the coupling constant dependence of $\tilde \chi$
when doing the coupling constant integration.  This is not an additional
approximation either in the RPA or for the approximate $\epsilon$'s we use
here.

In order to develop long-range functionals, one may substitute
approximations for the dielectric function based on the free electron gas
into \eref{eq:XACF_POL}.  To obtain tractable expressions it will normally
be necessary to make still further approximations. In this case it is
desirable to use the additional approximations only for the non-local part
of $E_\xc$ so as to avoid destroying the accuracy of the LDA in the
high-density regions. Ideally one would subtract from \eref{eq:XACF_POL}
the LDA version of the {\em same} approximation, and would envisage adding
back a better version of the LDA.  Here we make a similar, but more
tractable subtraction, which takes the form
\begin{equation}
E_\xc^0 = \iiabf{0}{\infty}{u}{2 \pi} \real{\trace{\log(\epsilon)}} -
E_{\rm self}.
\label{LDA}
\end{equation}
$E_\xc^0$ has the property that it is a good approximation for a slowly
varying system, becoming exact for a uniform system.  For density
variations slow on the scale of the range or width of
$\epsilon(\vv{r},\vv{r'})$, it agrees with the LDA, the trace in \eref{LDA}
replacing the integral over density.

Subtracting \eref{LDA} from \eref{eq:XACF_POL}, one obtains
\begin{equation}
\label{eq:XACF_NL}
E_\xc^\nl = \iiabf{0}{\infty}{u}{2 \pi} \real{\trace{\log(\epsilon^{-1}
\nabla\cdot\epsilon\cdot\nabla G)}}.
\end{equation}
We will call this the non-local exchange-correlation energy, although for
models more general than those used in this paper, an additional
short-range correction must be applied to make $E_\xc^0$ correspond
precisely to the LDA, and hence to make $E_\xc^\nl$ the deviation from the
LDA.

The approximations considered in this paper contain no non-local exchange
component, in effect making \eref{eq:XACF_NL} our approximation for the
non-local correlation energy $E_{c}^{nl}$.  Using this fact, together with
$\trace{\log x} = \log(\det{x})$ and $\nabla^2G=1$, we obtain
\begin{equation}
\label{eq:XACF_NL2}
E_c^\nl = \iiabf{0}{\infty}{u}{2 \pi} \log \norm{
\det(1+\epsilon^{-1}[\nabla,\epsilon]\cdot \nabla G)}
\end{equation}
where the notation $[A,B]$ means the commutator.  We will later use the
fact that \eref{eq:XACF_NL2} involves only the determinant to good
advantage.

\section{Method for Planar Geometries}

Now we are in a situation to discuss what $\epsilon$ to use.  We will in
this paper concentrate on the simple case of jellium systems. Our aim is to
find an \emph{efficient} way of exploring the planar translational
invariance of the jellium system -- not only to decrease the number of
spatial integrations.  The major difficulty in evaluating
\eref{eq:XACF_NL2} is the determinant, which is $\ordo{N^3}$ in the general
case, $N$ being the number of grid points in a discrete representation.
This holds true even in one dimension. So, instead of allowing a completely
general $\epsilon$, we aim at approximations resulting in
\emph{differential} operators only, for which the determinant is known to
be $\ordo{N}$, hence significantly simpler to calculate.

In the particular case of planar translational invariance, \ie for planar
surfaces or slabs, we use an approximate form that is made local in the
coordinate perpendicular to the surface,
\begin{equation}
\label{eq:ANSATZ}
\epsilon(\vv{r},\vv{r'}) = \delta(z-z') \int \frac{d^2k}{(2\pi)^2}
\epsilon_k(z) e^{i \vv{k} \cdot (\vv{r}-\vv{r'})},
\end{equation}
where $\vv{k}$ is a wave vector parallel to the surface.  Keeping the fully
non-local form along the symmetry plane allows, {\it e.g.}, the effect of
the cutoff, which was introduced artificially in previous approximations of
this type,\cite{AnRy97,HuRyLuLa98} to occur in a natural way laterally.

The first thing we note about \eref{eq:ANSATZ} is that we easily form the
inverse
\begin{equation}
\epsilon^{-1}(\vv{r},\vv{r'}) = \delta(z-z') \int \frac{d^2k}{(2\pi)^2}
\epsilon_k(z)^{-1} e^{i \vv{k} \cdot (\vv{r}-\vv{r'})}.
\end{equation}
Evaluating the commutator then yields
\begin{equation}
\epsilon^{-1} [\nabla,\epsilon] = \hat{z} \delta(z-z') \int
\frac{d^2k}{(2\pi)^2} \frac{\epsilon_k'(z)}{\epsilon_k(z)} e^{i \vv{k}
\cdot (\vv{r}-\vv{r'})},
\end{equation}
where the {\em prime} indicates differentiation with respect to $z$.  In
what follows we shall substitute $l(z)=\log(\epsilon(z))$, yielding
$l'(z)=\epsilon'(z)/\epsilon(z)$.

In the same basis we express the Green's function,
\begin{equation}
G(\vv{r}-\vv{r'}) = \int \frac{d^2k}{(2\pi)^2} G_{k}(z-z') e^{i \vv{k}
\cdot (\vv{r}-\vv{r'})},
\end{equation}
where
\begin{equation}
\label{eq:Gk}
G_k(z-z') = -\frac{1}{2 k} e^{-k \norm{z-z'}}.
\end{equation}
Since the logarithm of \eref{eq:XACF_NL} can be expanded in powers, the
integration over $k$ may be singled out, and we may express the non-local
correlation energy per surface area~($A$) as
\begin{equation}
\label{eq:XACF_POWER}
E_c^\nl/A = \iiabf{0}{\infty}{u}{2 \pi} \int \frac{d^2k}{(2\pi)^2} \log
\norm{ \det(1+\l_k' {\partial}_z G_{k})}.
\end{equation}

In \eref{eq:XACF_POWER}, the determinant is given in terms of integral
operators. To take advantage of the locality of the Laplacian, we use
$(\partial_z^2-k^2) G_k = 1$ to express it in terms of differential
operators
\begin{equation}
\label{eq:TRIDAG}
\det\left(1+l_k' \partial_z G_{k}\right) = \frac{\phi}{\phi_0},
\end{equation}
where
\begin{equation}
\phi= \det(\partial_z^2-k^2+l_k' \partial_z)
\label{eq:phidef}
\end{equation}
and where $\phi_0$ is the empty space ($\epsilon=1$) value of
\eref{eq:phidef}.  The step from \eref{eq:XACF_POWER} to \eref{eq:TRIDAG}
requires that the differential operators are defined throughout the whole
space.

Our ultra-fast method is made possible by the observation that the
determinants in \eref{eq:TRIDAG} can be written down, not only for the full
system, but also for a subdivision of it. Related determinants for the
subsystem satisfy a simple second-order differential equation as a function
of subsystem size.  Thus by a simple renormalization, one may evaluate
$E_c^\nl$ with the same effort as finding the charge induced by an applied
electric field.  A similar relation holds also in several dimensions, which
will be explored in another paper.

To make this more concrete, let us suppose that $\epsilon_k(z)$ varies only
in the interval $0<z<L$ (which will eventually be extended to infinity) and
takes the same value at either end point. This is the case for parallel
surfaces or slabs of identical materials. Then for each value of $z$ we can
define a determinant $\phi(z)$ for the subsystem extending from $0$ to $z$.
It is clear then from Eqs.~(\ref{eq:XACF_POWER}), (\ref{eq:TRIDAG}), and
(\ref{eq:phidef}) that $E_c^\nl$ is given by
\begin{equation}
\label{eq:XACF_phi}
E_c^\nl/A = \lim_{L \rightarrow \infty} \iiabf{0}{\infty}{u}{2 \pi} \int
\frac{d^2k}{(2\pi)^2} \log \frac{\phi(L)}{\phi_0(L)}.
\end{equation}
As discussed in Appendix A, the determinants $\phi(z)$ and $\phi_0(z)$
individually have oscillating signs that do not occur in their quotient.
However, the envelope determinants $\tilde{\phi}(z)$ and
$\tilde{\phi}_0(z)$ can be scaled so that they satisfy the simple
differential equation
\begin{equation}
(\epsilon_k \tilde\phi')' = k^2 \epsilon_k \tilde\phi,
\label{eq:diffeq}
\end{equation}
together with the boundary conditions that $\tilde\phi(0)=0$ and
$\tilde\phi(L) =1$.  In terms of $\tilde\phi$, we obtain
\begin{equation}
\label{eq:XACF_NICE}
E_c^\nl/A = - \lim_{L \rightarrow \infty} \iiabf{0}{\infty}{u}{2 \pi} \int
\frac{d^2k}{(2\pi)^2} \log \frac{\tilde\phi'(0)} {\tilde\phi'_0(0)},
\end{equation}
where the {\em prime} indicates differentiation with respect to $z$, which
in this case is the subsystem size. However, note that $\tilde\phi$ is also
just the electrostatic potential as a function of distance $z$, within a
system having a potential difference across it along with a prescribed
variation in $\epsilon$. Thus the calculation of the determinant becomes a
simple electrostatic problem which is easily solved.

To illustrate this, consider the case of two identical parallel surfaces a
distance $d$ apart, when $d$ is much larger than the thicknesses of the
surface-healing layers. Solving \eref{eq:diffeq} for the described boundary
conditions then becomes a trivial matching problem (see Appendix B), which
after insertion into \eref{eq:XACF_NICE} immediately leads to the Lifshitz
formula\cite{Lifshitz56,AnHuApLaLu98}
\begin{equation}
\label{eq:Lifsh}
E_c^\nl/A = \iiabf{0}{\infty}{u}{2 \pi} \int \frac{d^2k}{(2\pi)^2} \log
\norm{1-\rho^2 e^{-2 k d}} +2 \gamma_\nl.
\end{equation}
Here $\rho=(\epsilon_b-1)/(\epsilon_b+1)$, $\epsilon_b$ being the bulk
dielectric function, and $\gamma_\nl$ is defined by
\begin{equation}
\label{eq:gammanl}
\gamma_\nl = (E_c^\nl(d \rightarrow \infty)-E_c^\nl(0))/2A.
\end{equation}
Since, by construction, $E_c^\nl=0$ for a uniform ($d=0$) system,
$\gamma_\nl$ may equivalently be defined as the non-local correlation
contribution to the surface tension of a single surface.

The original ACF \eref{eq:XACF} is now reduced to a set of simple
electrostatic calculations, each one being an $\ordo{N}$ operation instead
of $\ordo{N^2}$, a major simplification. Of course the success of the
method depends on how well we can reproduce the true dielectric function
using our approximate form \eref{eq:ANSATZ}.  The only approximation made
so far is the assumption of a local dielectric function perpendicular to
the surface.

\section{Approximate Dielectric Function}

Equation (\ref{eq:XACF_phi}) or (\ref{eq:XACF_NICE}) provides the basis for
a functional that describes the van der Waals interaction between planar
objects. To turn these equations into density functionals, we have to
introduce quantities that depend on the density $n(\vv{r})$. Our suggestion
is based on an approximate dielectric function $\epsilon_k$ that depends on
the local density $n(\vv{r})$. It utilizes experiences from the homogeneous
electron gas and from experimental studies of the dynamical structure
factor $S(\vv{q},\omega) \propto
\imag{\frac{1}{\epsilon(\vv{q},\omega)}-1}$, where $\hbar \vv{q}$ and
$\hbar \omega$ are the momentum and energy losses, respectively, of a
photon or a charged particle being scattered while passing a bulk
sample. There is a peak in $S(\vv{q},\omega)$, the plasmon peak, sharp in
the ideal electron gas and of varying width in real materials.  This peak
carries most of the spectral strength and has $\omega$ equal to the plasma
frequency $\omega_p$ as $q \rightarrow 0$, and then a dispersion with a
limiting behavior $\omega_q \rightarrow q^2/2m$, the kinetic energy of one
electron, in the impulse-approximation, valid in the Compton-scattering
limit, $q \rightarrow \infty$.  In the electron-plasmon coupling one
focuses on the \emph{magnitude} and \emph{position} of the sharp plasmon
peak, and neglects the broadening, \ie $\epsilon$ is described in a
plasmon-pole approximation.\cite{Lu69} A dispersion law like
\begin{equation}
\label{plasmon_pole}
\omega_q^2 = \omega_p^2 + (v_F q)^2/3 + (q^2/2 m)^2
\end{equation}
has been shown to efficiently account for the average behavior of
plasmon-like excitations and for correlation properties of the homogeneous
electron gas.\cite{Lu69} Introducing the electron density via
$\omega_p^2(z) = 4 \pi n(z)$ and $v_F(z)=(3 \pi^2 n(z))^{1/3}$ in Hartree
units, $n(z)$ being the electron density profile in this planar case, the
dielectric function can be written as
\begin{equation}
\label{eq:epsilon}
\epsilon(z,k,iu) = 1+\frac{\omega_p^2(z)}{u^2+(v_F(z) q(k))^2/3+q^4(k)/4},
\end{equation}
where the imaginary frequency $\omega=iu$ is introduced.

Alternatively, \eref{eq:epsilon} may be viewed as an interpolation between
the exact small- and large-$q$ behavior of the Lindhard expression for the
frequency-dependent dielectric function.  However, we see little point in
using such an elaborate expression, since our concern here is to
investigate how well local approximations to the dielectric function work
in highly non-uniform systems.

In directions parallel to the surface, our approximation \eref{eq:epsilon}
allows fully for the non-diagonality of $\epsilon$ with respect to the
corresponding spatial coordinates, as implied by a Fourier transform with
respect to the parallel wave vector $k$. However, in directions
perpendicular to the surface, our approximation takes $\epsilon$ to be
diagonal in the coordinates $z$ and $z'$.  It is thus taken to be local not
only in this sense, but in the additional sense that it is a function only
of the local density. To compensate, we retain the corresponding component
$q_\perp$ of the wave vector $q$ in the right side of \eref{eq:epsilon} as
a {\it parameter}, so that everywhere $q^2 = k^2+q_\perp^2$.  We thus take
$1/q_\perp$ to be a constant measure of length over which $\epsilon$ is
effectively nonlocal. The dispersion perpendicular to the surface in
\eref{plasmon_pole} is in this way replaced by a parameter that we will fix
to some length scale appropriate for the surface.

For physical reasons such a length scale should be associated with
intrinsic electron-gas parameters like the screening length or the extent
of the correlation hole.  There are of course several choices available.
It must be kept in mind that we are after long range surface properties in
a variety of environments.  These properties are determined by various
response functions introduced by Feibelman,\cite{Fe80} of which the
simplest,
\begin{equation}
\label{eq:dimage}
d(iu)=\frac{\ii{z} z n_{\rm ind}(iu,z)}{\ii{z} n_{\rm ind}(iu,z)},
\label{eq:d_function}
\end{equation}
is the centroid of induced charge when a uniform electric field is applied
perpendicularly to the surface. However, for the van der Waals properties
of a planar surface, a related function $D(iu)$ as defined by Hult {\it et
al}.,\cite{HuAnLuLa96}
\begin{equation}
\label{eq:centroid}
D(iu)=\frac{\epsilon_b(iu)-1}{\epsilon_b(iu)+1}
\frac{\epsilon_b(iu)}{\epsilon_b(iu)+1} d(iu),
\end{equation}
is more important.\cite{ZaKo76,Feibelman82,HoAp84,PeZa84} In particular the
$D$-function arises in connection with the calculation of van der Waals
planes, which are determined not only by the $D$ of the surface in
question, but also by a response function of the {\it other} body. For
example, for a surface in the vicinity of an isotropic atom, the van der
Waals plane $Z$ is given by\cite{ZaKo76,HoAp84,PeZa84,HuAnLuLa96}
\begin{equation}
Z_\vdw= \frac{1}{4\pi C_3}\int_{-\infty}^\infty \! du \, \alpha(iu)D(iu),
\label{eq:surfatom}
\end{equation}
where $\alpha(iu)$ is the polarizability of the atom and $C_3$ is the
coefficient of the leading $1/z^3$ term in the asymptotic form of the
interaction energy.  In order for the surface calculation to scale
correctly for a wide variety of atoms with different $\alpha(iu)$'s, it is
obviously important for our approximation to well reproduce $D(iu)$.
Similarly, for two parallel surfaces labeled A and B, the van der Waals
plane for surface A is given by
\begin{equation}
Z_\vdw^A = \frac{1}{(4\pi)^2C_2} \int_{-\infty}^\infty \! du \,
\rho^B(iu)D^A(iu) + \Delta Z^A,
\label{eq:surfsurf}
\end{equation}
where $\rho^B(iu)$ is the long wavelength surface response function of
surface B, $\rho^B(iu)=(\epsilon^B_b(iu)-1)/(\epsilon^B_b(iu)+1)$, and
$C_2$ is the coefficient of the leading $1/z^2$ term in the asymptotic form
of the interaction energy.  The main difference in cases such as this one,
where two infinite bodies are involved, is that terms involving multiple
reflections no longer vanish in the asymptotic limit, and should be
explicitly included as indicated by the $\Delta Z^A$ term in
\eref{eq:surfsurf}.  However, it is common experience that in the
asymptotic limit multiple reflection terms are typically small enough to
treat in perturbation theory,\cite{Ha93,AnHuApLaLu98} a conclusion that we
agree with.  Therefore, the first term in \eref{eq:surfsurf} is dominant,
and the conclusion that $D(iu)$ is the key quantity to obtain accurately
remains.  Note, however, that in the numerical calculations presented
later, the full form of \eref{eq:surfsurf}, including all multiple
reflections (see Eqs.~(8) and (9) of \ocite{AnHuApLaLu98}), is used.

Thus we opt to choose $q_\perp$ based on the premise that the $D$-function
should be reproduced accurately.  We implement this by choosing $q_\perp$
so that $D(0)$ agrees exactly with full LDA calculations of this quantity,
a procedure precisely analogous to that of \ocite{HuRyLuLa98}.  In this
way, the parameter $q_\perp$ may be indirectly determined as a function of
the electron gas parameter $r_s$.  This determination gives $q_\perp\sim
1/r_s \sim k_{\rm F}$ as expected from previous arguments.

The constant $q_\perp$ smoothly limits the response at small $k$ and small
$u$ values. It thus replaces the sharp cutoff used in the earlier
scheme.\cite{HuRyLuLa98} In doing this, we consciously violate the Lifshitz
limit, obtaining somewhat smaller vdW coefficients than
\ocite{AnHuApLaLu98}, since the choice $q_\perp>0$ affects the $k
\rightarrow 0$ limit of \eref{eq:epsilon}.  This also means according to
Eqs. (\ref{eq:surfatom}) and (\ref{eq:surfsurf}) that the van der Waals
planes will be predicted to be somewhat too far from the surfaces in this
approximation.  The $d$-function, \ie \eref{eq:d_function} will also be too
large, significantly so at small $u$.

\begin{figure}[h!]
\centerline{\psfig{figure=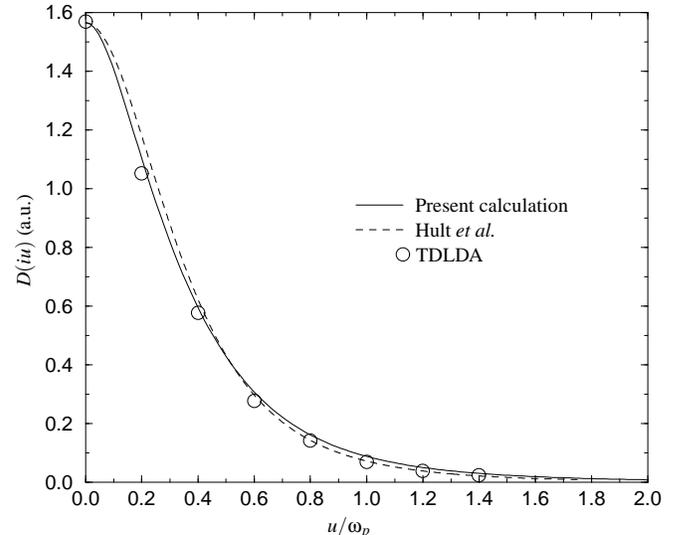,width=8.6cm}}
\caption{
\label{fig:Diu2}
The $D$-function $D(iu)$ for a jellium profile of $r_s=2$.  Solid line:
Present calculation.  Dotted line:
Reference\protect\onlinecite{HuAnLuLa96}.  Circles: $D$-function based on a
TDLDA calculation.\citep{Liebsch86} }
\end{figure}

\begin{figure}[h!]
\centerline{\psfig{figure=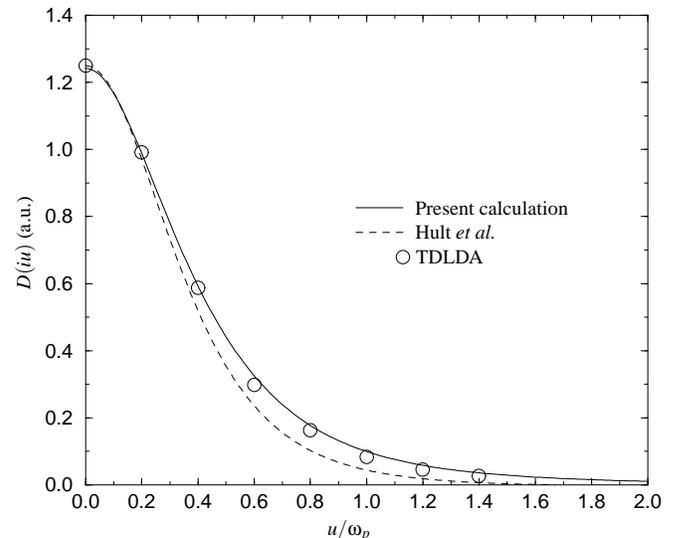,width=8.6cm}}
\caption{
\label{fig:Diu4}
The $D$-function $D(iu)$ for a jellium profile of $r_s=4$.  Solid line:
Present calculation.  Dotted line:
Reference\protect\onlinecite{HuAnLuLa96}.  Circles: $D$-function based on a
TDLDA calculation.\citep{Liebsch86} }
\end{figure}

Far more important, though, is the fact that our simple dielectric function
reproduces the dynamic properties of the $D$-function very well, as shown
in Figures \ref{fig:Diu2} and \ref{fig:Diu4}.  Thus the overall scaling
properties of our theory in a variety of non-uniform van der Waals
environments should continue to be correct.

In order to investigate the approximation with regard to key properties, we
use as input to our functional a set of self-consistent single-surface
jellium densities in the metallic range ($r_s=2-5$). The exact form of the
density profiles is however not that important; by construction, a
non-local energy is expected to depend less sensitively on the exact form
of the density, and results in this paper show that this is indeed the
case.

\begin{figure}[t!]
\centerline{\psfig{figure=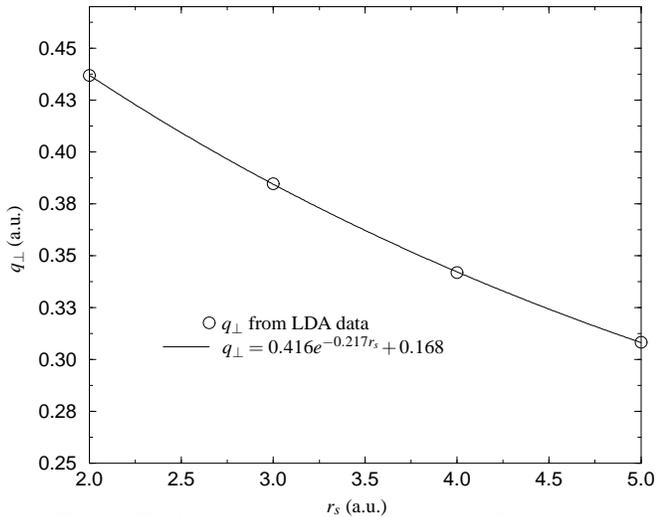,width=8.6cm}}
\caption{
\label{fig:qperp}
Optimal values of $q_\perp$, calculated for jellium surfaces of different
densities, described by $r_s$. Circles: Optimal $q_\perp$ as a function of
$r_s$, calculated from LDA data.\citep{SeSoGa86} Solid line:
Equation~(\ref{eq:interpol}).  }
\end{figure}

The optimal $q_\perp$ values found by fitting to the values of $D(0)$ of
\ocite{SeSoGa86} are shown in \fref{fig:qperp} as a function of the
electron-gas parameter $r_s$, well accounted for by the simple
interpolation formula
\begin{equation}
\label{eq:interpol}
q_\perp=0.416 e^{-0.217 r_s}+0.168.
\end{equation}
The variation is roughly 50 percent over the whole metallic range,
indicating a rather small overall sensitivity.

\tref{tab:centroid} shows how the quantities $D(0)$ and $Z_\vdw$ varies
with $r_s$, using the parameterization \eref{eq:interpol}.  In the fourth
and fifth columns, the values of the present method for the van der Waals
plane $Z_\vdw$ are compared to those of \ocite{AnHuApLaLu98}.  Although the
slightly larger coefficients are expected from considerations earlier in
this section, the fact that the numbers come out almost \emph{independent}
of $r_s$ is somewhat of a surprise.
\begin{table}
\caption{
\label{tab:centroid}
Static and dynamic image-plane positions (a.u.) of the jellium surface,
calculated as a function of $r_s$.  }
\begin{tabular}{ddddd}
$r_s$ & $D(0)$\tablenotemark[1] & $D(0)$\tablenotemark[2] &
$Z_\vdw$\tablenotemark[3] & $Z_\vdw$\tablenotemark[4] \\ \tableline 2.00 &
1.57 & 1.57 & 1.15 & 0.96 \\ 2.07 & 1.55 & & 1.15 & \\ 2.30 & 1.48 & & 1.15
& \\ 2.66 & 1.41 & & 1.14 & \\ 3.00 & 1.35 & 1.35 & 1.13 & 0.87 \\ 3.28 &
1.32 & & 1.14 & \\ 4.00 & 1.24 & 1.25 & 1.12 & 0.79 \\ 5.00 & 1.17 & 1.17 &
1.09 &
\end{tabular}
\tntext{1}{This work.}  \tntext{2}{Reference\protect\onlinecite{SeSoGa86}.}
\tntext{3}{This work.}
\tntext{4}{Reference\protect\onlinecite{AnHuApLaLu98}.}
\end{table}

\section{Results}
\begin{figure}[b!]
\centerline{\psfig{figure=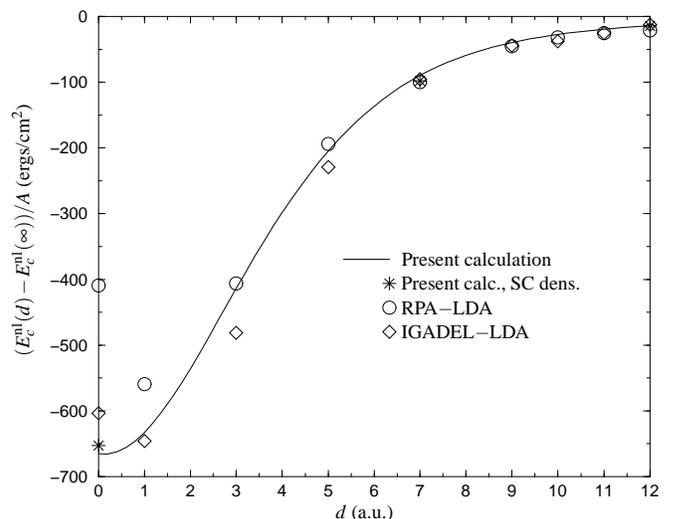,width=8.6cm}}
\caption{
\label{fig:dobtest}
Small-separation ($d$) variation of the non-local correlation-interaction
energy (ergs/cm$^2$) between two parallel jellium slabs of $r_s=2.07$ and
width $5$~a.u.\citep{DoWa98} Solid line: Present calculation.  Stars: Same
calculation but using the self-consistent densities of \ocite{DoWa98}.
Circles: RPA$-$LDA of \ocite{DoWa98}.  Diamonds: IGADEL$-$LDA of
\ocite{DoWa98}.  }
\end{figure}

To test our approximate DF, we solve \eref{eq:diffeq}, using
\eref{eq:epsilon}, and insert the result into \eref{eq:XACF_NICE} for a
known system, consisting of two parallel jellium slabs, separated by a
distance $d$. Figure \ref{fig:dobtest} shows results in ergs/cm$^2$ (${\rm
ergs}/{\rm cm}^2 = 0.6423$ $\mu$Ha$/a_0^2$) for the non-local
correlation-interaction energy per surface area, $(E_{\rm c}^\nl(d)-E_{\rm
c}^\nl(\infty))/A$. The results using the $r_s=2.07$ value of
\eref{eq:interpol} are compared to those of a recent RPA calculation of a
slab system of $r_s=2.07$ by Dobson and Wang,\cite{DoWa98} as well as to
those of their approximate DF (IGADEL).  The saturation effects are found
to be substantial in this small-separation region, and we judge all the
proposed density functionals in the figure to give good accounts of the
non-local correlation energy.  The agreement with IGADEL reflects the
inherent similarities between IGADEL and our approximation. The
tractability of the latter is however not reflected in the table, but has
to be stated here (about a thousand times faster than IGADEL for this
particular system, due to the overall lower computational complexity of our
DF), together with the claim that this gives great prospects for a
tractable future general DF.

The calculation presented in \fref{fig:dobtest} is performed using a simple
superposition of the densities of two separate slabs (obtained from $d=12$
data) as input. The difference between the non-local correlation energy for
the self-consistent density of the slab system and that obtained by
superposition turns out to be very small, as indicated by the stars in
\fref{fig:dobtest}.

The minor deviations at $d$ close to zero between our results (solid curve)
and those of the full RPA calculation (circles) are due to our somewhat
crude treatment of the width of the exchange-correlation hole perpendicular
to the surface, a small price one has to pay for a tractable DF.

Another important property is the surface correlation energy.  This
quantity has recently been calculated for a jellium surface within the
RPA,\cite{PiEg98} an approximation thought to give the long range
correlation effects accurately.  The long-range part ($\gamma_{\rm nl}$)
may be extracted using the data of Kurth and Perdew,\cite{PeKu99} by
subtracting the LDA contribution to the same quantity.  In
\tref{tab:zerogamma}, the result (column 3) is compared with our
approximation (column 2), calculated as the non-local correlation energy
given by Eqs. (\ref{eq:XACF_NICE}) and (\ref{eq:epsilon}), for
single-surface jellium at various values of $r_s$, using
\eref{eq:interpol}. We note that the two approximations differ on average
by only $13$ percent.  This is perhaps the strongest indication that indeed
the physical approximations made in this paper are robust.
\begin{table}
\caption{
\label{tab:zerogamma}
Non-local correlation contribution to the surface energy of the jellium
surface, $\gamma_\nl$ (ergs/cm$^2$), as function of $r_s$, calculated from
Eqs. (\ref{eq:XACF_NICE}) and (\ref{eq:epsilon}) together with
\eref{eq:interpol} and self-consistent single-surface LDA densities.
Comparison is made with results for $\gamma_\nl$ obtained by subtracting
the LDA contribution to the surface correlation energy\citep{PeKu99} from
results of self-consistent RPA calculations for a jellium
surface.\citep{PiEg98} }
\begin{tabular}{ddd}
$r_s$ & $\gamma_\nl$\tablenotemark[1] & $\gamma_\nl$\tablenotemark[2] \\
\tableline 2.00 & 451 & 476 \\ 2.07 & 408 & 435 \\ 2.30 & 300 & 332 \\ 2.66
& 196 & 226 \\ 3.00 & 138 & 163 \\ 3.28 & 106 & 129 \\ 4.00 & 60 & 74 \\
5.00 & 31 & 39
\end{tabular}
\small{ \tntext{1}{Non-local correlation energy for single-surface
jellium.}\\ \tntext{2}{RPA$-$LDA (in the RPA) of
Refs.\protect\onlinecite{PiEg98} and\protect\onlinecite{PeKu99}.}  }
\end{table}

A remarkable fact already indicated in \fref{fig:dobtest} is that the
non-local correlation energy is quite insensitive to the exact form of the
density profile. To further test this assumption, we use a linear
superposition of two identical self-consistent LDA single-surface densities
(SLDA) to calculate the non-local correlation surface energy according to
\eref{eq:gammanl}.  The result closely follows the column 2 result of
\tref{tab:zerogamma}, with a mean error of only 3 percent.  This
observation adds to the accumulated findings supporting the use of
superpositions of single-fragment densities in a future general DF.

\begin{figure}
\centerline{\psfig{figure=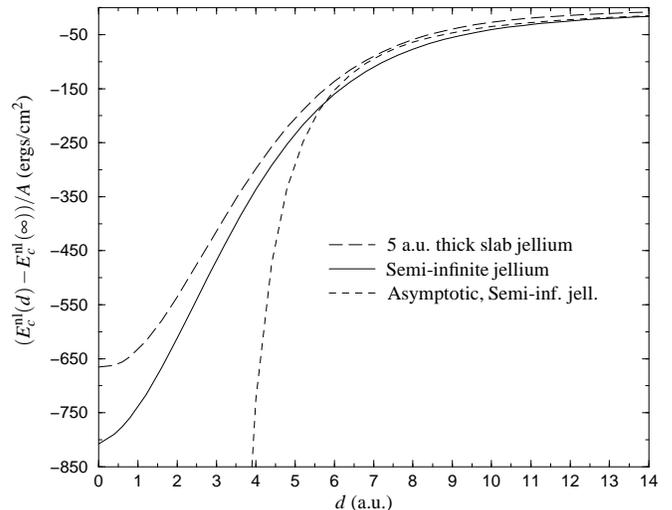,width=8.6cm}}
\caption{
\label{fig:rs207}
Small-separation ($d$) variation of the non-local correlation-interaction
energy (ergs/cm$^2$) for two semi-infinite jellium surfaces of $r_s=2.07$
(SLDA density; solid curve), calculated from our DF defined by Eqs.
(\ref{eq:XACF_NICE}), (\ref{eq:epsilon}), and (\ref{eq:interpol}).  The
results are compared to the corresponding interaction between two
$5$~a.u. slabs with edge separation $d$ (same as in \fref{fig:dobtest};
dashed curve).  The dotted curve shows the asymptotic form of the
interaction, $E = -C_2/(d-2 Z_\vdw)^2$, with $C_2=1.34$~$\mu$Ha and
$Z_\vdw=1.15$~a.u., calculated as in \ocite{AnHuApLaLu98} but with the
dielectric function \eref{eq:epsilon}.  }
\end{figure}

\begin{figure}[t!]
\centerline{\psfig{figure=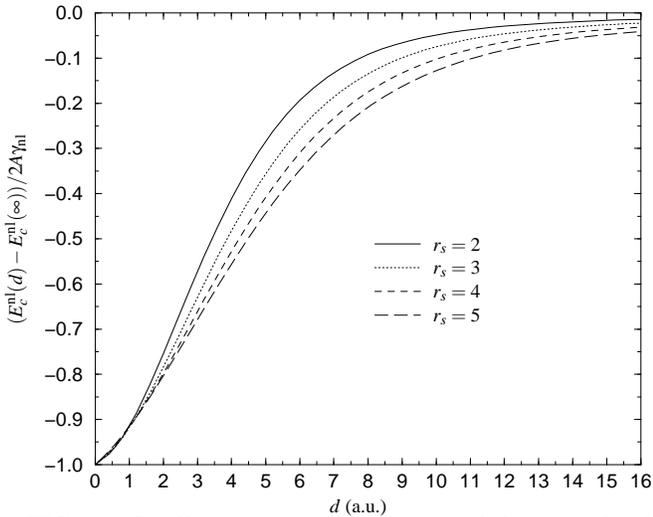,width=8.6cm}}
\caption{
\label{fig:scale}
Small-separation ($d$) variation of the normalized non-local
correlation-interaction energy, $(E_{\rm c}^\nl(d)-E_{\rm
c}^\nl(\infty))/2A\gamma_\nl$, for a range of $r_s$ values. The
$\gamma_\nl$ values are presented in \tref{tab:zerogamma}.  }
\end{figure}

After these successful tests of the predictive power of the DF, defined by
Eqs. (\ref{eq:XACF_NICE}), (\ref{eq:epsilon}), and (\ref{eq:interpol}),
applications to other systems where no other results are available might be
done with confidence.  Here we present results of an application to two
semi-infinite jellia of identical $r_s$, having their parallel surfaces a
distance $d$ apart (Figs. \ref{fig:rs207} and \ref{fig:scale}).  The input
densities are obtained by linear superposition of two self-consistent
single-surface LDA densities (SLDA).  We stress that our DF is very fast;
obtaining a single value for a given density profile and a given separation
only takes a few seconds on a typical workstation of today.

Figure \ref{fig:rs207} shows the variation of the calculated non-local
correlation-interaction energy for two semi-infinite jellia of $r_s=2.07$
as a function of the separation $d$. It illustrates two facts: the
significant deviation from the corresponding quantity for two thin jellium
slabs (same as in \fref{fig:dobtest}; $5$~a.u. wide) and the substantial
saturation effects, the latter by comparing with the results of the
asymptotic DF formula,\cite{AnHuApLaLu98} applying the dielectric
function~\eref{eq:epsilon}.

In \fref{fig:scale}, we present the normalized non-local
correlation-interaction energy $(E(d)-E(\infty))/2A\gamma_\nl$ in the
metallic range, showing how the interaction varies with $r_s$. The curves
depend only weakly on $r_s$ when scaled in this manner.

\section{Concluding remarks}

In summary, we have studied the basis for a DF accounting for vdW
interactions, by starting in the manner of our previous work but with
essential generalizations to small separations between interacting
objects. A systematic approach for the construction of such a DF is
described, together with a very efficient method to calculate the resulting
expressions. In the case of flat surfaces, results for the interaction of
two parallel jellium slabs are shown to agree with those of a recent RPA
calculation,\cite{DoWa98} and we show that input densities can be
successfully approximated by a superposition of the electron densities of
the interacting fragments. Results for the surface energy of jellium are
compared favorably with other studies.\cite{PeKu99} As a prediction of the
theory, the interaction energy between two parallel jellia is calculated
for all separations $d$ and in the whole metallic range.  The well known
asymptotic behavior ($E \propto 1/z^2$) is obtained for large $d$, and as
$d$ becomes smaller, substantial saturation effects are predicted.

The major significance of these results is the demonstration that such
numbers can be calculated accurately at a reduced computational complexity
and hence greatly improved speed. We have shown that for a subclass of
dielectric functions, the resulting expressions for the non-local
correlation energy may be calculated very efficiently, and that even a
simple approximation to the dielectric function yields valuable insight and
reproduces several physical properties of flat surface and slab
models. Furthermore, we have indicated that generalizations to
three-dimensional systems are possible, and that the results here suggest
such an attempt to be a fruitful one. In this way there should be a basis
for applications to numerous physical, chemical, and biological systems,
such as vdW bonds in crystals, liquids, adhesion, soft condensed matter
(e.g., biomacromolecules, biosurfaces, polymers, and membranes), and
scanning-force microscopy.

\acknowledgments We thank J.~Dobson for providing us with several numerical
density profiles obtained in \ocite{DoWa98} for parallel slabs, and
J.~Perdew for providing computer code to generate density profiles for
isolated jellium surfaces.  Work at Rutgers supported in part by NSF Grant
DMR 97-08499.  Financial support from the Swedish Natural Science Research
Council and the Swedish Foundation for Strategical Research through
Materials Consortium no. 9 is also acknowledged.

\appendix

\section{Details of the evaluation of the determinants $\phi$ and $\phi_0$}
Here we show how ratios of determinants like Eq. (\ref{eq:XACF_POWER}) can
be efficiently evaluated. To eliminate the oscillating sign problem
mentioned in the main text, we need to first go to a discrete
representation for the operator $(\partial_z^2-k^2+l_k' \partial_z)$
occurring in \eref{eq:phidef}.  Specifically we take $N$ points between $0$
and $L$ (for the full determinant), $n$ points between $0$ and $z$ (for
subsystem determinants), so that $z=(n/N)L$, with the spacing between
points $h=L/N$.  We use a similar representation for the empty space
operator $(\partial_z^2-k^2)$.  Thus, replacing the subsystem determinant
$\phi(z)$ by the discretisized $\phi_n$, we have
\begin{equation}
\phi_n = \det\M{ccccc}{ a_1 & b_1 & 0 & \ldots & 0 \\ c_1 & a_2 & \ddots &
\ddots & \vdots \\ 0 & \ddots & \ddots & b_{n-2} & 0 \\ \vdots & \ddots &
c_{n-2} & a_{n-1} & b_{n-1} \\ 0 & \ldots & 0 & c_{n-1} & a_n },
\end{equation}
where
\begin{eqnarray}
\nonumber a_n &=& -2-h^2 k^2 \\ \nonumber b_n &=& 1+\frac{h}{2} l_{k,n}' \\
c_n &=& 1-\frac{h}{2} l_{k,n+1}',
\label{eq:abcdef}
\end{eqnarray}
along with a similar relation for the empty space determinant $\phi_{0,n}$.
Being tridiagonal, the determinant can be evaluated in $\ordo{N}$ time, as
contrasted to the $\ordo{N^3}$ time applicable for a general determinant.

The determinant is readily expanded by minors, giving the recursion formula
\begin{equation}
\phi_{n+1} = a_{n+1} \phi_n - b_n c_n \phi_{n-1},
\label{eq:phidiscrete}
\end{equation}
with $\phi_0=1$ and $\phi_1=a_1$. It is clear from (\ref{eq:phidiscrete})
and (\ref{eq:abcdef}) that $\phi_n$ oscillates in sign from order to order,
an oscillation that is cancelled in the ratio \eref{eq:TRIDAG} by a similar
oscillation in $\phi_0$.

However, the envelope determinant $(-1)^n\phi_n$ satisfies a simple
differential equation in the continuum limit.  To make the exact form of
this differential equation identical to \eref{eq:diffeq}, we further scale
the envelope determinant $\tilde\phi_n$ through
\begin{equation}
\phi_n\equiv(-1)^nS_n\tilde\phi_n,
\label{eq:tildephi}
\end{equation}
where the scaling function $S_n\equiv\Pi_{i=1}^nb_i$ satisfies
\begin{equation}
S_n=b_n S_{n-1}
\label{eq:Sdiscrete}
\end{equation}
with $S_0=1$. Using (\ref{eq:abcdef}), we see that in the continuum
$h\rightarrow 0$ limit, (\ref{eq:Sdiscrete}) becomes
\begin{equation}
\frac{dS(z)}{dz}=\frac{1}{2} l_k'(z) S(z),
\label{eq:Scontinuous}
\end{equation}
with the boundary condition $S(0)=1$.  Equation~(\ref{eq:Scontinuous}) has
the solution
\begin{equation}
S(z)=\left(\frac{\epsilon_k(z)}{\epsilon_k(0)}\right)^{\frac{1}{2}}.
\label{eq:Ssolution}
\end{equation}
This means that if $\epsilon_k(L)=\epsilon_k(0)$, then the scaling due to
$S$ has no effect on the final result.  Otherwise, the argument of the
logarithm in \eref{eq:XACF_NICE} should be multiplied by $S(L)$ obtained
from (\ref{eq:Ssolution}).

The difference equation for $\tilde\phi$ is obtained by use of
(\ref{eq:tildephi}) and (\ref{eq:Sdiscrete}) in (\ref{eq:phidiscrete}),
yielding
\begin{equation}
b_{n+1}\tilde\phi_{n+1}+a_{n+1}\tilde\phi_n +c_n\tilde\phi_{n-1}=0,
\label{eq:phidif}
\end{equation}
with
\begin{equation}
\tilde\phi_0=1
\label{eq:boundarycondition1}
\end{equation}
and
\begin{equation}
\tilde\phi_1-\tilde\phi_0=-(1+a_1/b_1).
\label{eq:boundarycondition2}
\end{equation}
Substitution of (\ref{eq:abcdef}) into (\ref{eq:phidif}) yields
\begin{eqnarray}
&&(\tilde\phi_{n+1} - 2\tilde\phi_n +\tilde\phi_{n-1})
-h^2k^2\tilde\phi_n\nonumber\\ &&+\frac{h}{2} l'_{k,n+1}
(\tilde\phi_{n+1}-\tilde\phi_{n-1})=0.
\label{eq:tpde}
\end{eqnarray}
In the continuum ($h\rightarrow 0$) limit, this becomes
\begin{equation}
\tilde\phi''(z) -k^2\tilde\phi(z) + l_k'(z)\tilde\phi'(z)=0.
\end{equation}
This equation is the same as \eref{eq:diffeq} in the main text, although
there we used the notation $\tilde\phi$ for a particular solution, while in
this appendix it represents the general solution.

This general solution to \eref{eq:diffeq} can be written in the form
\begin{equation}
\tilde\phi(z)=\alpha\tilde\phi_\alpha(z)+\beta\tilde\phi_\beta(z)
\end{equation}
where $\tilde\phi_\alpha(0)=1$ and $\tilde\phi_\alpha(L)=0$, while
$\tilde\phi_\beta(0)=0$ and $\tilde\phi_\beta(L)=1$.  The boundary
condition (\ref{eq:boundarycondition1}) implies that $\alpha=1$, while
(\ref{eq:boundarycondition2}) combined with (\ref{eq:abcdef}) implies that
$\tilde\phi'(0)=1/h$ as $h$ approaches zero.  This means that
$\tilde\phi_\alpha$ can be neglected, since $\beta$ will be of order $1/h$.
Specifically we have $\tilde\phi'(0)\rightarrow\beta\phi'_\beta(0)=1/h$, so
that
\begin{equation}
\tilde\phi(L)\equiv\beta=\frac{1}{h\tilde\phi'_\beta(0)}.
\label{eq:tpl}
\end{equation}
The large coefficient ($\propto 1/h$) cancels out in the ratio of
(\ref{eq:tpl}) and the analogous free-space expression, so we may write
\begin{equation}
\frac{\phi(L)}{\phi_0(L)}=\frac{\tilde\phi(L)}{\tilde\phi_0(L)}S(L)
=\frac{\tilde\phi'_{0,\beta}(0)}{\tilde\phi'_\beta(0)}S(L),
\label{eq:proportions}
\end{equation}
where the continuum version of (\ref{eq:tildephi}) was used to obtain the
first identity, after noting that the oscillating signs of the numerator
and denominator cancel in the ratio, and that the free space value of $S$
is unity.  Finally substituting this into \eref{eq:XACF_phi}, we obtain,
using (\ref{eq:Ssolution}),
\begin{equation}
E_c^\nl = - \iiabf{0}{\infty}{u}{2 \pi} \int \frac{d^2k}{(2\pi)^2} \log
\frac{\tilde\phi'_\beta(0)\sqrt{\epsilon_k(L)}}
{\tilde\phi'_{0,\beta}(0)\sqrt{\epsilon_k(0)}}.
\end{equation}
This is our most general result, which reduces to \eref{eq:XACF_NICE} when
$\epsilon_k(0)=\epsilon_k(L)$.  Note that in the main text, we used the
notation $\tilde\phi$ and $\tilde\phi_0$ for the particular solutions that
are called $\tilde\phi_\beta$ and $\tilde\phi_{0,\beta}$ here.

\section{Details on the Lifshitz limit}
Let $\epsilon_k(z)=1$ in between two surfaces a distance $d$ apart, and let
$\epsilon_k(z)=\epsilon_b$ in the bulk of each surface.  Furthermore, let
$d$ be large enough so we can assume sharp boundaries between the three
regions. Then, the interaction energy $(E_c^\nl(d)-E_c^\nl(\infty))/A$ may
be calculated exactly.  Note that although the interaction energy may be
calculated this way, the constant component contributing to the surface
energy may not, but needs a much more involved calculation.

The matching problem becomes solving \eref{eq:diffeq} for $\phi$ and
$\phi_0$ in the regions left bulk (lb), middle region (m), and right bulk
(rb).  Let $\phi_0=e^{k x}$, with the origin lying on the boundary between
the left and middle regions.  Now we want to find the $\phi$ that
approaches zero far into the left bulk, and $e^{k x}$ far into the right
bulk. In the left bulk, we must have
\begin{equation}
\phi_{\rm lb}=a e^{k x},
\end{equation}
$a$ being an arbitrary constant. Note that the wanted quantity
$\phi_0'(0)/\phi'(0)$ equals $1/a$.  On the boundary between the left bulk
and the middle region, $\phi$ must be continuous, and $\phi'$ must have a
discontinuous step with the size of $\epsilon_b$, reflecting the fact that
the displacement field $\epsilon_b \phi'$ must also be continuous.  The
solution in the middle region now becomes
\begin{equation}
\phi_{\rm m}=a (\cosh(k z)+ \epsilon_b \sinh(k z)).
\end{equation}
The same matching conditions apply between the middle region and the right
bulk, although here, we only need to consider the solution growing to the
right, since we want to match $\phi$ to the value of $\phi_0$ infinitely
far into the right bulk. Obeying the matching conditions means the solution
in the right bulk becomes
\begin{equation}
\label{cgrowing}
\phi_{\rm rb}^{\rm growing}=\frac{\phi_{\rm m}(d) k+\phi'_{\rm
m}(d)/\epsilon_b}{2 k} e^{k (x-d)}.
\end{equation}
Equating $\phi_0$ with (\ref{cgrowing}) determines the coefficient a, and
the solution becomes
\begin{equation}
\label{eqxclif}
\phi_0'(0)/\phi'(0) = \frac{\phi_{\rm m}(d) k+\phi'_{\rm
m}(d)/\epsilon_b}{2 k a e^{k d}} = \frac{1-\rho^2 e^{-2 k d}}{1-\rho^2},
\end{equation}
with $\rho=(\epsilon_b-1)/(\epsilon_b+1)$.  It is clear from
(\ref{eqxclif}) that the constant contribution to
$\log(\phi_0'(0)/\phi'(0))$ and hence to the surface energy is given by
$-\log(1-\rho^2)$. As discussed earlier in this appendix, that is not the
correct constant and should be excluded, yielding \eref{eq:Lifsh}.


\begin{thebibliography}{10}
\bibitem{FootNotation} The notation $A B$ for the operation $\int d^3r''
A(\vv{r},\vv{r}'') B(\vv{r}'',\vv{r}')$ is used throughout the text, Tr
means the trace, and $\delta^3(\vv{r}-\vv{r}')$ is denoted by $1$ where
appropriate.
\end{thebibliography}


\end{document}